# Cavity Optocapillaries

Shai Maayani, Leopoldo L. Martin, Samuel Kaminski and Tal Carmon.

*Technion - Israel Institute of Technology, 32000 Haifa, Israel*

**Droplets, particularly water, are abundant in nature and artificial systems. Thermal fluctuations imply that droplet interfaces behave like a stormy sea at the sub-nanometer scale. Thermal capillary-waves have been widely studied since 1908 and are of key importance in surface science. Here we use an optical mode of a µdroplet to probe its radius fluctuation. Our droplet benefits from a finesse of 520 that accordingly boosts its sensitivity in recording Brownian capillaries at 100-kHz rates and $1\pm0.025$ ångström amplitudes, in agreement with natural-frequency calculation and the equipartition theorem. A fall in the fluctuation spectrum is measured below cutoff at the drop's lowest-eigenfrequency. Our device facilitates resonantly-enhanced optocapillary-interactions that might enable optical excitation (/cooling) of capillary droplet-modes, including with the most-common and important liquid - water.**

Single sentence: We activate a µdroplet as a hybrid optocapillary resonator, and access its thermal capillaries while benefiting from resonantly-enhanced sensitivity.

Thermal capillary waves [1-6] were first suggested by Smoluchowski [7] and are most important in processes such as spontaneous rupture of liquid-films [8, 9] and droplet coalescence [10]. Thermal capillaries were measured via deviations in the law of reflection from liquid interfaces. For example, light [1] and X ray [2-5] scattering techniques from liquid interfaces revealed details on the thermal capillary roughness and dynamics as well as on the reduction of the surface tension in short length scales [5]. This reduction is significant to the theory of interfaces [11, 12]. Recently, thermal capillary waves and droplet coalescence were directly visually observed in a colloidal suspension [6] while exploiting scaling up of the lengths when going from molecules to mesoscopic colloidal-particles (diameter ~100 nm). These colloids were used for measuring thermal capillaries since direct tracking of the dynamics of individual atoms is not experimentally feasible. In a droplet, thermal fluctuations were measured by interfering the reflection from its interface with a reference beam [13].

Generally speaking, extending interference experiments as in reference [13] to measure interference between many reflections from the liquid interface (instead of one) can provide an interface-displacement sensitivity that linearly increases with the number of reflections. In fact, placing two mirrors facing each other to constitute an optical resonator can provide the mirror displacement with a resolution as fine as the optical wavelength divided by the number of roundtrips [14]. This number of optical roundtrips is therefore called "finesse"; basically, because it describes how fine a resolution this cavity can provide. Indeed, optical cavities are commonly used to boost the sensitivity in displacement measurement devices, including gravitational wave detectors [15]. Relevant to our experiment, light can also resonate while circulating on the circumference of a droplet. Continuous research in droplet optical resonators [16-20] has led to the recent demonstration of water droplets with ultrahigh-finesse [21]. In this demonstration, photons are confined in a tight, 180 nm region, adjacent to the droplet interface (Fig. 1A inset) where capillaries reside. For this reason, it is natural to ask if droplets with high optical-finesse can provide resonantly-enhanced access to thermal capillaries. Ideally, a $10^6$-finesse droplet, as reported in [21], can improve displacement resolution by a factor of a million. Yet being modest here, we investigate a 520-finesse system, where a 1% change in optical transmission represents a 0.026 Å droplet deformation.

Beyond allowing a finesse-scaled boost in resolving capillaries, a resonator co-hosting light and capillaries can pave the way for new frontiers. That is, and from a broader-impact point of view, opto-capillary cavities can support a controlled energy-exchange between light and capillaries. Such an energy exchange is possible, for instance, by operating at the resonance sideband region. As an example, operating at the blue (/red) side of the optical resonance can facilitate optical excitation (/cooling) of capillary oscillations, which we believe will be soon possible with our optocapillary resonator.

In the following, we discuss the droplet from a "capillary" point of view. Droplets represent a fundamental structure of self-contained liquid, bounded almost completely by free surfaces. Unlike free propagating capillary waves with continuous frequency spectrum, droplets host a set of discrete thermal capillary resonances that dominate the droplet spectrum at the relatively low damping (by viscosity) region. The lowest capillary eigenfrequency is near the frequency of a capillary wave whose length is comparable with the droplet size, and known as cutoff frequency.

To couple between the drop's optical and capillary oscillations, we park our laser wavelength near an optical resonance. As one can see in Fig. 1B, transmission fluctuation, $\Delta T$, will scale with radius fluctuation, $\Delta r$, [14] as given by

$$(1) \quad \Delta T = \frac{Q}{r} \Delta r \ ,$$

where Q is the optical quality factor of the droplets and r is its radius. A system with input optical frequency near the steepest region of the Lorentzian-shaped resonance can therefore translate radius fluctuation to changes in the optical transmission through the cavity. This transmission can be easily measured with a photodetector. Relying on this principle, we use the optical mode [16-21] as a probe that optically interrogate the drop's thermal capillary oscillations [22, 23]. As one can see in Fig. 1B, a spherical droplet can capillarlly oscillate. The eigenfrequency of this shape oscillation [22, 23] is

$$(2) \quad f = \sqrt{2\gamma/(\pi^2 \rho r^3)} \cong \sqrt{k/(\tfrac{2}{3}\pi r^3 \rho)} \; ,$$

where $\rho$, $\gamma$ are the liquid density and surface tension at the air-liquid interface. Using this analytical solution, we estimate the spring constant of this system, $k$, assuming that the effective vibrating mass is half the droplet mass. The thermal capillary fluctuation, Δr, in this droplet is calculated using

$$(3) \quad k_B T = k \left\langle \Delta r^2 \right\rangle \; ,$$

where T is the temperature and $k_B$ is the Boltzmann constant. Substituting typical values, the thermal capillary motion of an r=10 μm droplet, at room temperature, is Δr = 1.8 Å.

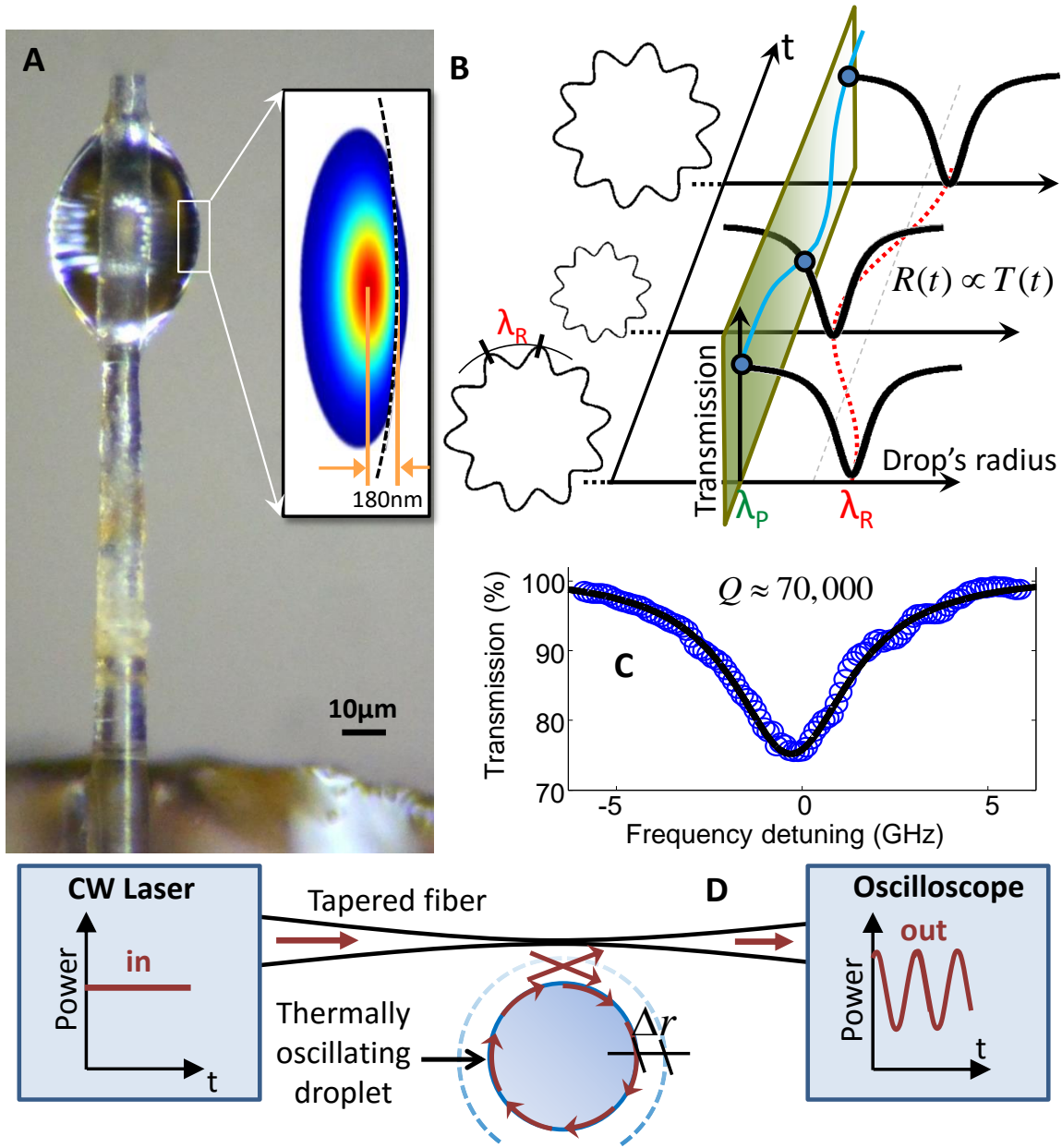

**Figure 1: Experimental setup:** (A) Micrograph of the water μdroplet along with its calculated fundamental optical mode shape (inset). (B) A Lorentzian transfer function (black) converts radius drifts (red) into changes in optical transmission (blue). (C) Monitoring droplet resonator transmission while scanning a wavelength through resonance reveals its optical quality factor. (D) Experimental setup scheme. Laser wavelength is 980 nm.

We *fabricate* our water-droplet resonator by dipping one side of a silica stem in a practically unlimited water reservoir [21]. As silica is hydrophilic, water creeps up the stem forming a droplet at its end (Fig. 1A). We then use a tapered fiber to evanescently couple light into the water droplet [24]. These tapers are convenient for coupling light in and out of solid and liquid [18, 19] resonators since they end with a standard optical fiber at both their input and output sides. Next, we *characterize* the optical resonance of the droplet by scanning the pump-laser wavelength through a resonance (Fig. 1C). We calculated an optical quality of 70,000 from the resonance linewidth.

Monitoring transmission near a resonance, such as in Fig. 1B, can teach us about liquid droplet's size fluctuation. This technique was used to similarly interrogate shape oscillations in solid devices [25]. We should note here that the resonance wavelength (Fig. 1B) does not instantly track radius fluctuations; rather is delayed by the cavity photon-lifetime. Still, as photon lifetime in our 70,000 Q experiment is 70 ps, we well oversample [26] our ~100 kHz capillary fluctuations.

We *experimentally* record the thermal capillaries of our water droplet, by tuning our laser wavelength near resonance (as illustrated in Fig. 1c) and then *measuring* radius fluctuation by monitoring transmission. Figure 2 shows the measured drop's fluctuations that reveal, as expected [27], quasi sinusoidal oscillations in the temporal domain. This type of quasi-sinusoidal oscillation was theoretical predicted [27] to be irregular over a period of time comparable to the ripplon lifetime. As the irregularity period here (Fig. 2A) is longer than the oscillation cycle, we can say that these droplet fluctuations are at the underdamped regime. This claim is in agreement with our calculations using the relatively low-viscosity of water [28].

Upon examination of the oscillation in the frequency domain (Fig. 2 B-C), we focus our analysis on two peaks that dominate over their vicinity. Performing a finite element calculation (ANSYS Fluent™) reveals droplet eigenmodes (Fig. 2 B-C) with eigenfrequencies near the rates of the experimentally measured peaks. As expected, the lowest-frequency mode relates to motion along the axial direction, which is longest in our system. Similarly, the higher-frequency mode (Fig 2B right) involves motion along the shorter radial-direction.

In contrast to free propagating capillary waves where all frequencies are allowed, a droplet cannot host modes with wavelengths larger than the droplet diameter.  Indeed and as seen in Fig. 2A, we observe a reduction in thermal capillaries below the first mode (near 50 kHz) where the density of capillary-states drops. Such a reduction is typically called cutoff and the eigenfrequency of the lowest state is called the cutoff frequency.

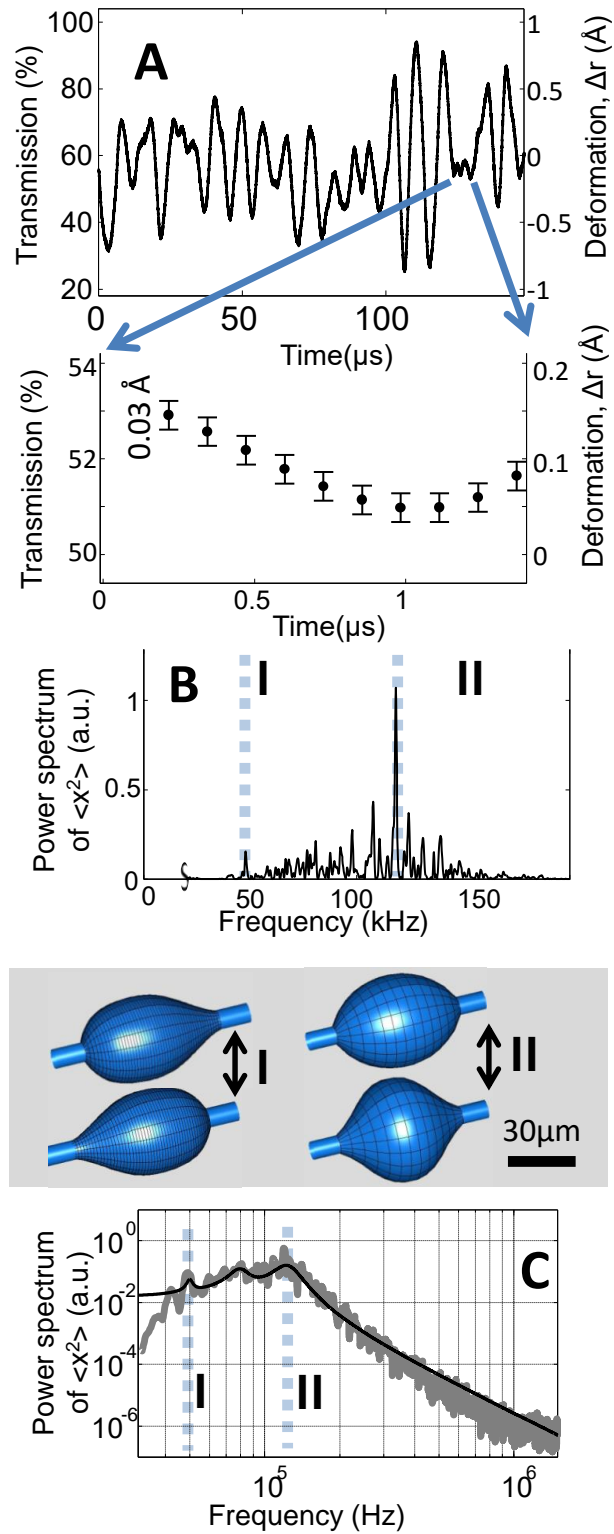

**Figure 2, Experimental results: Droplet fluctuations** are shown in the temporal (A) and frequency domains (B) where the oscillation frequencies that dominate over their vicinity are linked to calculated droplets modes (Supplementary movie 1-2) (inset). The blue arrows lead to a zoom-in plot where each of the points represents the transmission average over a 125 ns period. The error bar represents the standard deviation for each averaging. Presenting the capillary spectrum on a log-log scale (C) reveals a drop at high frequencies where the black line represents a Lorentzian fit. The Lorentzian 'skirt' dominates at the high frequencies at the right hand side of (C). The stem holding the droplets has a radius of 5 µm and the droplet radius is 16 µm and the optical wavelength. Deformation in the 3D plots is exaggerated in the graphics.

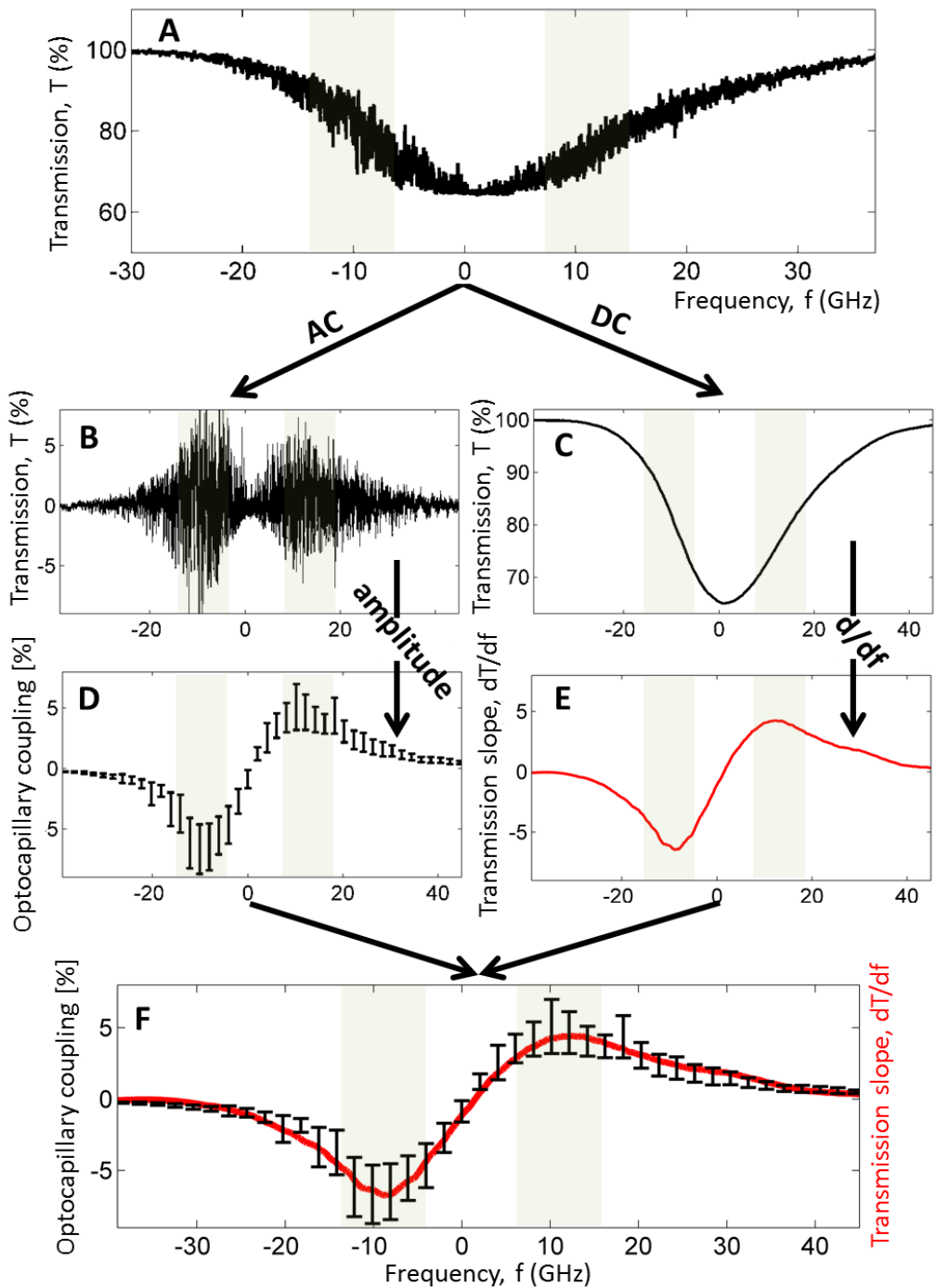

**Figure 3, Experimentally reaching maximal optocapillary coupling** at the region where the optical resonance has the steepest transmission slope. Slowly scanning the laser frequency

while light is coupled to the optical resonance reveals its spectral transmission (A). Spectral transmission is then split to AC components (B) representing transmission oscillations (by capillaries) and to DC (C) representing the Lorentzian shape of the optical resonance. The optocapillary coupling is measured through the amplitude of the AC transmission oscillation (D). As one can see in (F), optocapillary coupling (F, black) is maximal at the green background regions (in A-F) where cavity transmission is the steepest (E). Points in (D) represent averaging over oscillation amplitude (at this region) and the error bars represent the standard deviation of the oscillation amplitude. The diameter of the droplet is 118 µm and its optical quality factor is 19500.

Looking at Fig. 1B, one expects that the highest detection sensitivity for thermal capillaries is at the steepest slope region of the Lorentzian-shape. Contrary to that region, the Lorentzian center exhibits a zero first-derivative where sensitivity for detecting thermal-capillaries vanishes. To find that optocapillary coupling indeed scales with the first derivative of the Lorentzian representing optical transmission, we measure transmission oscillation again, but now scanning our laser frequency through resonance. We use here an octane droplet submerged in water [29]. This octane-core water-clad droplet is almost perfectly spherical, and its position (in respect with the tapered-fiber coupler) is controlled using optical tweezers, as explained in [29].

As one can see in Fig. 3, we scan our laser frequency through the regions of maximal- and minimal-optocapillary coupling. Of course, the regions of maximal optocapillary coupling (Fig. 3 A-F green background) are interesting from the practical point of view of needing high sensitivity. The region of zero optocapillary coupling (that appears between the green backgrounds) is less practical for sensing capillaries; yet, it depicts the Lorentzian optocapillary transfer function at its singular flat-top region. Capillaries at this region do not affect the cavity optical transmission. We end our experiment here by showing that optocapillary coupling indeed scales with the slope of the optical-resonance transmission (Fig. 3F) as expected from the illustration of optocapillary coupling (Fig. 1B). The fact that transmission oscillation follows the first derivative of (DC) transmission (Fig. 3F) makes it unlikely that oscillation originates from oscillation in coupling strength or in laser power. These are independent of the transmission slope. As for possible oscillation in laser frequency, we choose a laser with a specified frequency drift (NewFocus Velocity Laser TLB-6700) that is 4500 times smaller than the frequency drifts caused by the droplet capillaries. We experimentally confirm this laser spec by performing a control-group experiment to test the laser, as shown in Fig. 1D, against a solid silica "droplet" (silica stiffness is compared to that of steel) with an optical quality factor of 300 million.

In conclusion, we fabricate an optocapillary resonator; then use its optical resonance to probe thermal capillaries with resolution enhanced by its 520 finesse. We measure thermal capillary oscillations in a droplet as well as their cutoff with a resolution of 0.03 Å. Our technique is relatively simple as it is relies on regular water, uses standard optical fibers and photodetectors, operates on an individual droplet at room temperature and pressure, and works with no controller or feedback loop.

Energy exchange and interaction between light and sound (optoacoustics) or light and light (nonlinear optics) has been widely studied. Its most important enabler is a resonator with optical modes that overlap the wave they interact with. On the contrary, optocapillary resonators were rarely available. Light was therefore thought of as minimally interacting with capillaries. We believe that our optocapillary cavity will serve as a bridge between capillary and optical waves, and will enable a new type of experiments including in search for capillary rogue waves.